# Comprehensive fabrication of SNAP microresonators by a femtosecond laser

Qi Yu, Zhen Zhang, and Xuewen Shu *

*Wuhan National Laboratory for Optoelectronics and School of Optical and Electronic Information, Huazhong University of Science and Technology, Wuhan 430074, China*
*xshu@hust.edu.cn*

**Abstract** Surface nanoscale axial photonics (SNAP) microresonators with nanoscale effective radius variation (ERV) along optical fiber axis can be fabricated by inscribing axially oriented lines inside the fiber with a femtosecond laser. The optimization of variable parameters in the femtosecond laser inscription technique is of great significance for flexible and ultra-high precision control of ERV, which is vital to the performance of SNAP devices. Here, we present the first systematical investigation on the relationships between the various controllable fabrication parameters and the introduced ERV of the SNAP microresonators. Specifically, both the qualitative and quantitative processing principles were revealed. As a proof-of-principle, by comprehensively optimizing the fabrication parameters, we realized a SNAP microresonator with the characteristics of both small axial size and maximal ERV, which is almost impossible to realize with other SNAP fabrication techniques. Our work promotes the fs laser inscription technology to be a flexible and versatile approach for fabricating the SNAP devices with ultra-high precision, ultra-low loss and high robustness.

## 1. Introduction

Surface nanoscale axial photonics (SNAP), as an emerging ultra-low loss and ultra-precise whispering gallery mode (WGM) resonant circuit at the surface of an optical fiber, has many important potential applications [1]. By modifying a uniform fiber with nanoscale effective radius variations (ERV), a SNAP bottle microresonator enables the support of WGMs circulating near the fiber surface and slowly propagating along its axis [2]. Multifunctional SNAP devices have been proposed through engineering the properties of the WGMs which are controlled by the introduced ERV along the fiber axis, such as optical delay lines [3], optical buffers [4], slow-to-tunneling propagation crossover [5] and so on. In order to introduce and control the ERV of the optical fiber, many fabrication techniques have been proposed, such as the annealing of the focused $CO_2$ laser beams [6], femtosecond (fs) laser inscription [7], fiber tapering [8], strong bending of optical fibers [9] and internal ohmic heating [10]. To ensure the desired performances of SNAP microresonators, the high precision and flexibility are two key requirements of the SNAP fabrication techniques, which are remained challenges for researchers.

Recently, fs laser inscription technology [11-13] begins to play a significant role in fabricating SNAP devices. With the high peak power and short duration of fs pulses, the modified zone induced by the interactions between the fs pulses and transparent materials can be confined in a micro-sized focal volume [14,15], which makes the fs laser inscription have great advantages including ultra-precision, ultra-low loss and much smaller axial dimensions for the fabrication of SNAP microresonators compared with other techniques [6,8-10]. As we know, rectangular SNAP microresonators with high steepness along the fiber axis can be easily and precisely fabricated by inscribing two cross-sections in the optical fiber with focused fs laser pluses [16]. SNAP microresonators can be induced with high repeatability and high precision by fs laser post-processing [17]. Although the fs laser has been demonstrated as a powerful tool to fabricate and modify the SNAP microresonators, there is no systematical research on the relationships between the various controllable parameters of the fs laser inscription technique and the introduced ERV, which is crucial for promoting the versatility of the fs laser fabrication technique in SNAP platform. In addition, systematically investigating the fs laser fabrication technology on SNAP microresonators can help one to understand the underlying physical mechanisms of the interactions between fs pulses and transparent materials.

Here, aiming to solve the above problem, we investigated the relationships between the ERV of the SNAP microresonators and the controllable fabrication parameters of the in-fiber axial inscription approach with a fs laser, including the number of the axially inscribed lines, the spacing between the adjacent axially inscribed lines, the single pulse energy, the spacing between the adjacent inscribed layers, the number of inscribed layers, and the ratio of the repetition rate and the translating speed. Both the qualitative and quantitative processing principles were revealed based on the control variate method. Then, by comprehensively optimizing the fabrication parameters, we realized a SNAP microresonator with the characteristics of both small axial size and maximal ERV, which is almost impossible to fabricate with other SNAP fabrication techniques. Our work gives better understanding on the interactions between the fs pulse and silica, and promotes the fs laser inscription technology to be a flexible and versatile approach for fabricating the SNAP devices with ultra-high precision and ultra-low losses.

## 2.  Experimental setup and Theoretical background

Our experimental setup is illustrated in Fig.1. The ultra-short pulses ($\lambda = 520$ nm) with a duration of 220 fs were used to inscribe axially oriented lines inside optical fibers with the cladding radius $r_0 = 40\ \mu m$. The laser beam was focused by an oiled objective lens (Olympus UMPLFL 63×) with a numerical aperture (NA) of 1.4. Each fiber was stripped from its protective coating and immersed in refractive-index-matching oil to compensate for the lens effect of the curved fiber surface. A high resolution translation stage was used to translate the fiber relative to the laser beam with a desired speed.

By inscribing axially oriented lines inside the optical fiber, the refractive index change and the induced stress in the modification volume result in the increase of fiber radius along the fiber circumference, thus forming a SNAP microresnonator [7], which is schematically shown in Fig. 1(b). This process can be modeled using a thick-walled cylinder by approximating the axial inscription as an axially symmetric cylinder having the effective radius $r_i$. By assuming the modification region $r < r_i$ with the characteristic radius $r_i$ pressurize the remaining part of the fiber $r_i < r < r_0$, the variation of the fiber radius can be obtained from the Lame's equation [18]:

$$\Delta r = \frac{2 r_i^2 P_i}{E r_0}, \qquad (1)$$

where $E = 76\ GPa$ is the Young modulus of the silica, and $P_i$ is the pressure applied to the internal surface of the cylinder $r_i < r < r_0$.

The illustration in Fig. 1(a) is the schematic diagram of the inscription area. The modification volume is formed by several inscribed layers parallel to each other and to the fiber axis, which are symmetrically distributed along the $y$ axis with the spacing between two adjacent layers of $\Delta y$. Each layer is made up of several parallel inscribed lines along the fiber axis with a length of $L_z$. The distance between two adjacent lines is $\Delta x$. Thus, $(\Delta x, \Delta y, L_z)$ is the characteristic geometric parameters of the modification volume.

As we know, the spectrogram of the SNAP microresonator, which is the surface plot of WGMs transmission spectra $T(\lambda, z)$, can be utilized to characterize the introduced ERV along the fiber axis. The transmission spectra $T(\lambda, z)$ is measured as a function of wavelength $\lambda$ and coordinate $z$ along the SNAP microresonator axis using an input–output microfiber. The taper of the microfiber is coupled with the SNAP fiber to excite the WGMs propagating along the

SNAP fiber with a small propagation constant $\beta_c(\lambda, z)$ which vanishes at the cutoff wavelengths (CWs) $\lambda_c(z)$ [1,2]. The introduced ERV $\Delta r(z)$ is usually resolved by the maximum variation of the WGM cutoff wavelength $\Delta\lambda_c(z)$ as $\Delta r(z) = \Delta\lambda_c(z) r_0 / \lambda_c$ [1]. In this paper, for convenience, we use the cutoff wavelength variation (CWV) to substitute the ERV calculation because the CWV can be easily obtained in the measured spectrum.

In our experiments, a biconical taper with a micron-diameter waist was contacted perpendicularly with the SNAP microresonator and translated along the SNAP fiber with a fixed step. The transmission light was detected by an optical spectrum analyzer (OSA) with a spectral resolution of 0.01nm, which is shown in Fig.1(b). The transmission spectrum at each position is normalized by removing the initial taper loss (spectral average of 0.2 dB) so that the transmitted power is 0 dB (no loss) in the absence of coupling.

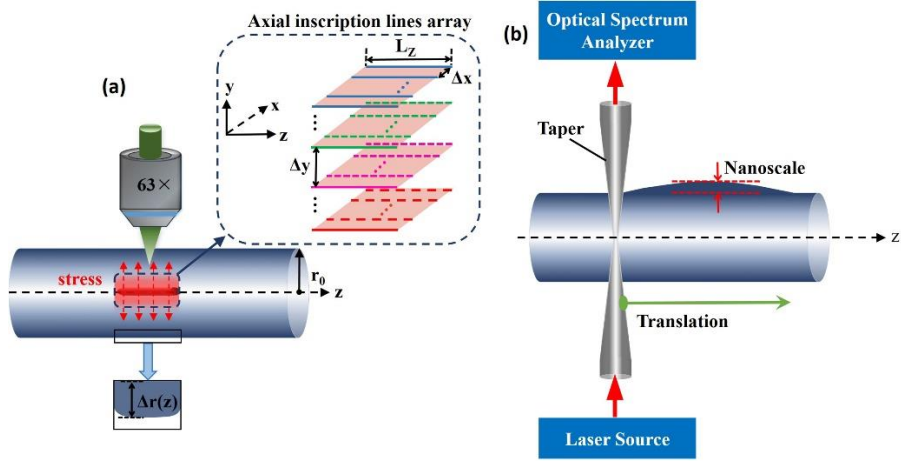

Fig. 1(a). Illustration of the fabrication setup of the nanoscale ERV introduced by the axially oriented lines inscription by a fs laser. (b) Illustration of the measurement setup of the SNAP microresonator.

## 3. Experimental results and discussions

The ERV introduced by the fs laser inscription is usually controlled by the following seven controllable parameters: the number of the axially inscribed lines $N$, the spacing between the adjacent lines $\Delta x$, the single pulse energy $E_s$, the spacing between the adjacent inscribed layers $\Delta y$, the number of inscribed layers $N_L$, and the ratio of the repetition rate of the ultra-fast pulses $f$ and the translating speed $v$. To obtain the quantitative relationships between the ERV(or CWV) and these parameters, the control variate method is used in our research.

### 3.1 The number of the axially inscribed lines

Firstly, in order to investigate the effect of the number of the inscribed lines on the introduced ERV, 16 SNAP microresonators were fabricated along the fiber with different $N$. During the fabrication process, we fixed the setting of the fabrication parameters $(L_z, E_s, \Delta x, N_L, \Delta y, f, v)$ to be (150 $\mu$m, 70 nJ, 3 $\mu$m, 1, 0 $\mu$m, 200 kHz, 20 $\mu$m/s), while only changed the parameter $N$ from 3 to 18 for 16 SNAP microresonators, respectively. Here, the number of lines are limited by the fiber radius and the spacing of adjacent lines, which ensures the inscribed regions inside are usually separated from the fiber surface by more than 10 $\mu$m to avoid introducing additional losses of microresonators [7]. The transmission spectrogram of the SNAP was measured with the axial spatial resolution of 2 $\mu$m, as presented in Fig. 2(a). We can get from Fig. 2(a) that the

CWV of each SNAP microresonator increases from 0.047 nm to 0.247 nm as the $N$ increased from 3 to 18. The response of the CWV of each SNAP microresonator to different number of the inscribed lines $N$ is linearly fitted with the slope of 0.0138 nm per line, as shown in blue line in Fig.2(b), which corresponds to the linear response of ERV to the number of the inscribed lines $N$ with the slope of 0.356 nm per line. Though the slope of the curve of ERV-to-N can be changed by adjusting the setting of the parameters $(L_z, E_s, \Delta x, N_L, \Delta y, f, v)$, the quasi-linear relationship is unchanged through our extensive experimental verifications.

From the experimental results above, we can get that the ERV is linearly proportional to the number of lines $N$ within a certain range. Therefore, it is suitable for us to consider the variation of the fiber radius introduced by multiple parallel inscribed lines as the accumulation of multiple independent thick-walled cylinders, thus the expression in Eq. (1) can be modified as

$$\Delta r = \sum_{k=1}^{N} \frac{2r_i^2 P_i}{E r_0} = N \frac{2r_i^2 P_i}{E r_0} . \qquad (2)$$

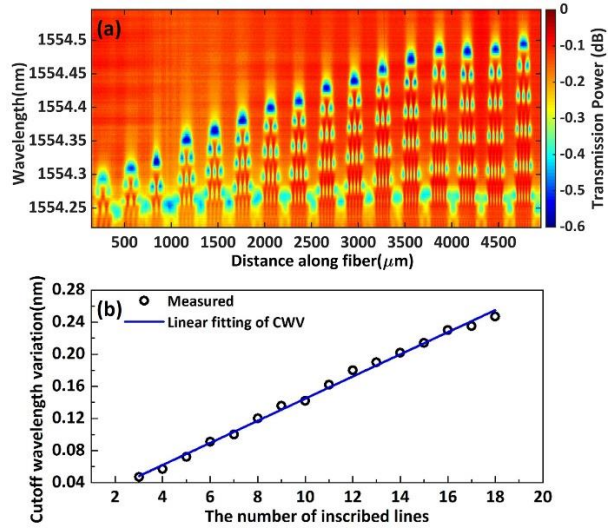

Fig.2 (a) Spectrogram of SNAP microresonators fabricated by different number of the inscribed lines varied from 3 to 18 respectively (left to right). (b) Linear response of the CWV to the number of the inscribed lines with the fitting of CWV: $\Delta \lambda_c(N) = 0.0138 \cdot N + 0.0068$.

*3.2 The spacing between the adjacent axially inscribed lines*

From Eq. (1), we know that the ERV is also proportional to the cross section area $\pi r_i^2$ of the inscribed line, which is geometrically affected by the spacing of the adjacent lines in case of overlapping. To investigate the relationship between the ERV and the line spacing $\Delta x$ inscribed in the same horizontal layer, we fabricated 6 groups of SNAP microresnators with the line spacing $\Delta x$ varied from 2 μm to 7 μm, respectively. Each group consists of 6 SNAP structures fabricated with different single pulse energy $E_s$ varied from 37 nJ to 110 nJ, respectively. Meanwhile, the other fabrication parameters $(L_z, N, N_L, \Delta y, f, v)$ were set as (150 μm, 5, 1, 0 μm, 200 kHz, 20 μm/s). The spectrograms of the SNAP microresonators are shown in Fig.3. The CWV of each SNAP microresonator with the fixed spacing between adjacent lines increases linearly with the increasing of $E_s$. However, the CWV of the SNAP microresonator with the same $E_s$ has no obvious linear relationship with the line spacing $\Delta x$, which can be seen in Fig.4(a). In each curve with the same single pulse energy, the CWV increases monotonically with the increasing of the line spacing, accompanying with the saturation feature at large line spacing. To explain this, we measured the width of each inscribed line fabricated with different single pulse energy under the microscope, which is shown in Fig.4(b). We found that each inscribed line using different single pulse energy has a different line width $r_i$, which determines an optimal spacing between adjacent lines to avoid overlaps. The overlaps may affect the accumulation of the total cross section area and thus the total ERV by all inscribed lines. To further explain this, we numerically calculated the CWV versus the line spacing with different line width by using Eq. (2). The theoretical curves are plotted with the dotted dash lines in Fig.4(a). If the line spacing $\Delta x$ is smaller than the line width $r_i$, the adjacent lines will overlap each other, which results in the total cross section area of all inscribed lines equal to the linear combination of the cross section area of each line minus the overlapped areas of all adjacent lines. According to Eq. (2), we can get that the CWV with the overlapped line spacing inscription is smaller than that with non-overlapped line spacing. If the line spacing $\Delta x$ is larger than the inscribed line width $r_i$, there is no overlap between the adjacent lines and the total cross section area is the linear combination of each line, thus the CWV (or ERV) reaches the maximal value under certain single pulse energy, which is dependent on the number of the lines and independent of the line spacing. It is seen from Fig. 4(a) that our experimental results are basically in agreement with the numerical calculation.

As an example, we demonstrated a clearly comparison of two typical microscope images in Fig. 4(a1) and (a2). Since the inscribed line width $r_i$ with $E_s$ of 110 nJ is 4.6 μm (see Fig. 4(b6)), the inscribed lines with the line spacing $\Delta x$ of 4 μm overlap each other (see Fig. 4(a1)), and the corresponding CWV of the SNAP microresonator is ~ 0.14 nm, which is smaller than that fabricated with the non-overlapped line spacing $\Delta x$ of 6 μm (see Fig. 4(a2)). Besides, based on the measured results of all line widths fabricated with different single pulse energy in Fig.4 (b), we suggest that the minimal line spacing in our non-overlapped fabrication can be set as 3 μm, 3 μm, 4 μm, 5 μm, 5 μm, 5 μm, respectively, marked in black circles in Fig.4(a). Overall, different singe pulse energy will induce different line width. So, we should ensure the two adjacent lines have the optimal line spacing larger than the line width of each line to avoid overlap, which is important for non-overlapped inscription to high-precisely control the ERV of SNAP microresonators.

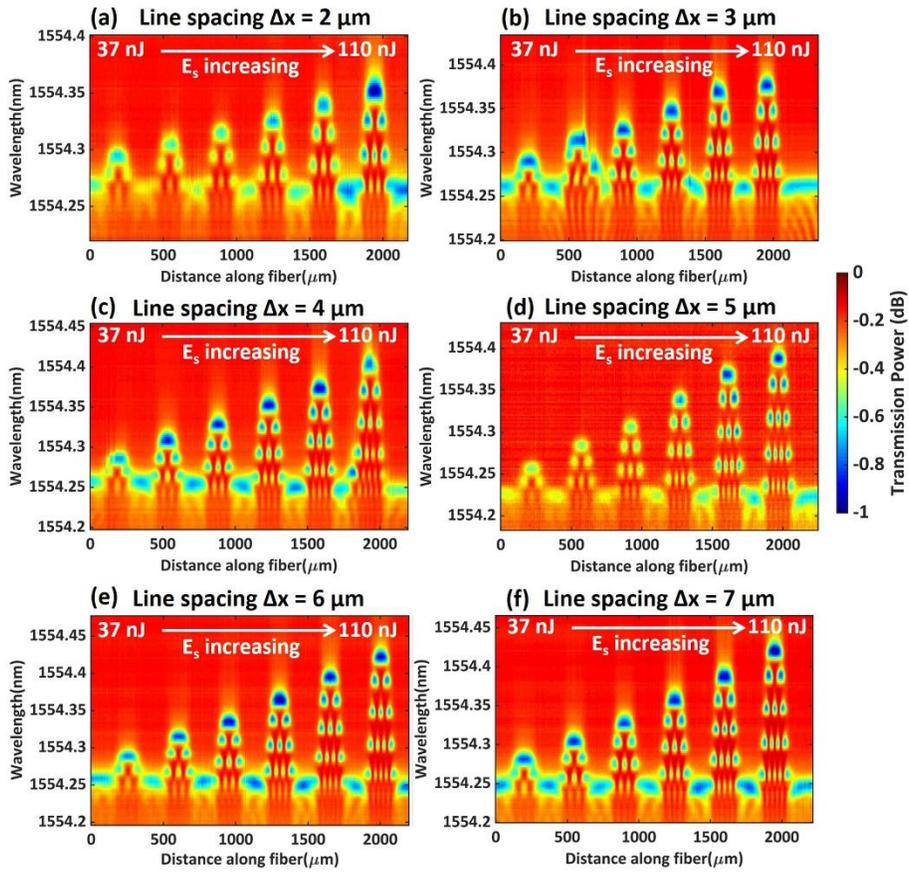

Fig.3 (a) Spectrogram of SNAP microresonators fabricated with the line spacing of (a) 2 $\mu$m, (b) 3 $\mu$m, (c) 4 $\mu$m, (d) 5 $\mu$m, (e) 6 $\mu$m, (f) 7 $\mu$m. Each group consists of 6 SNAP microresonators fabricated with the same parameters but different fabricated single pulse energy with 37 nJ, 50 nJ, 64 nJ, 79 nJ, 94 nJ, and 110 nJ from left to right, respectively.

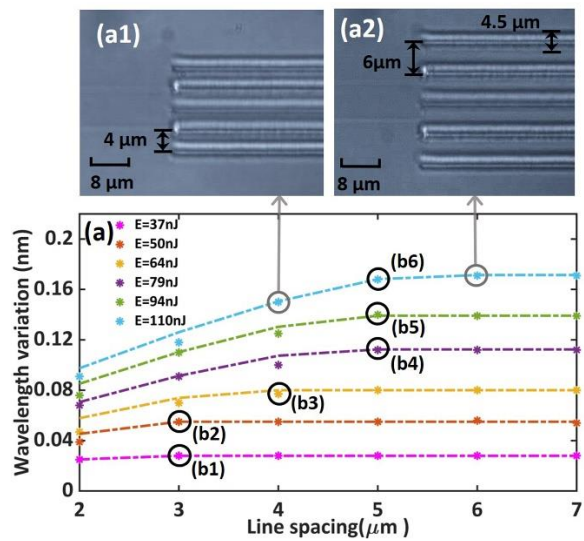

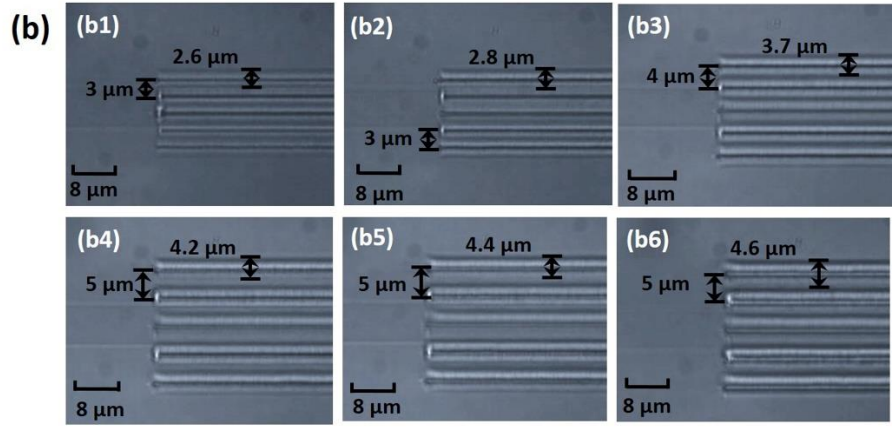

Fig.4 (a) The response of the CWV to the line spacing under different fabricated single pulse energy. The dotted dash lines depict the theoretical CWV based on Eq. (2), while the star points are experimentally measured. The microscope images of the line with (a1) 4 $\mu m$ and (a2) 6 $\mu m$, which are both fabricated with the same single pulse energy of 110 nJ. (b) Microscopes images of inscribed lines fabricated with different single pulse energy of (b1) 37 nJ, (b2) 50 nJ, (b3) 64 nJ, (b4) 79 nJ, (b5) 94 nJ, and (b6) 110 nJ, respectively.

### *3.3 The single pulse energy*

In section 3.2, we find that the single pulse energy $E_s$ can affect the ERV, which is critical for the design of the line spacing. To investigate the quantitative relationship between the single pulse energy and the ERV, we fabricated 12 SNAP microresonators along the fiber with different $E_s$ varied from 30 nJ to 110 nJ, respectively. During the fabrication process, other fabrication parameters $(L_z, N, \Delta x, N_L, \Delta y, f, v)$ were set to be (150 $\mu m$, 5, 5 $\mu m$, 1, 0 $\mu m$, 200 kHz, 20 $\mu m/s$). Note that, we set the line spacing $\Delta x$ of 5 $\mu m$ to avoid the overlap of the adjacent lines. The maximal single pulse energy is limited by the additional loss brought by the oil ablation outside the fiber surface under the high single pulse energy inscription. Figure 5(a) is the spectrogram of the SNAP microresonators with the axial spatial resolution of 4 $\mu m$. We can get that the CWV of each SNAP microresonator increases from 0.014 nm to 0.158 nm as $E_s$ increased from 30 nJ to 110 nJ. The response of the CWV of each SNAP microresonator to $E_s$ is linearly fitted with the slope of 0.0018 nm/nJ, as shown in pink line in Fig.5(b), which corresponds to the linear response of ERV to $E_s$ with the slope of 0.046 nm/nJ. The results indicate that the linear relationship between the ERV and the single pulse energy could enable the flexible and accurate fabrication of SNAP microresonators with sub-angstrom precision.

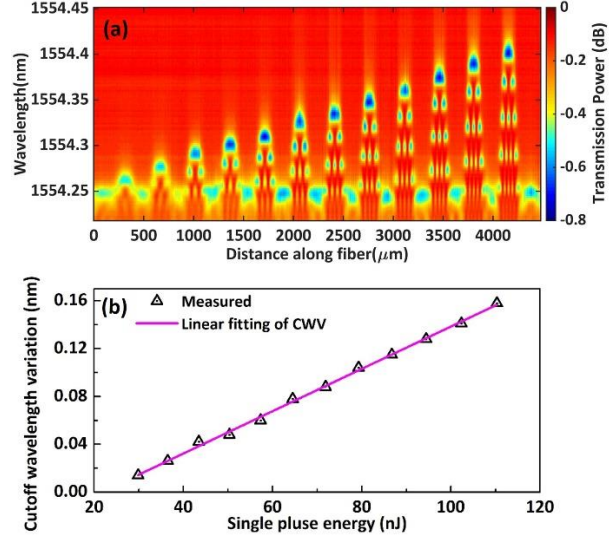

Fig.5 (a) Spectrogram of SNAP microresonators fabricated by different single pulse energy increasing from 30 nJ to 110 nJ (from left to right). (b) Linear response of the CWV to the single pulse energy with the fitting of CWV: $\Delta\lambda_c(E) = 0.0018 \cdot E - 0.0385$.

### 3.4 The number of the inscribed layers

To avoid introducing scratch at the fiber surface and additional losses of microresonators, the inscribed regions inside are usually separated from the fiber surface by more than 10 $\mu$m. This leads to the limitation of the number of the inscribed lines by the fiber radius and the line spacing in one inscribed layer. While in many cases, in order to introduce larger ERV with much more number of lines, we can also inscribe lines in the cladding area along the $y$ axis (see inset part of Fig.1(a)) with the spacing of adjacent inscribed layers $\Delta y$. It is curious for us to investigate the relationship between the ERV and the distance of the inscribed layer away from the central plane of fiber core, which could also help to quantify the ERV response to the number of the inscribed layers. Firstly, we fabricated 12 SNAP microresonators with different inscribed layer-to-core distances. For each SNAP microresonator, we inscribed two layers parallelly and symmetrically distributed around the fiber-core axis. We fixed the fabrication parameters $(L_z, E_s, N, \Delta x, N_L, f, v)$ of the 12 SNAP microresonators to be (150 $\mu$m, 50 nJ, 10, 3 $\mu$m, 2, 200 kHz, 20 $\mu$m/s), while only adjusted the inscribed layer-to-core distance from 2 $\mu$m to 24 $\mu$m with a step of 2 $\mu$m, respectively. Figure 6(a) is the spectrogram of the SNAP microresonators with the axial spatial resolution of 4 $\mu$m. The CWV of each SNAP microresonator decreases from 0.207 nm to 0.043 nm with the increasing of the layer-to-core distance from 2 $\mu$m to 24 $\mu$m. The CWV of the SNAP microresonator versus the layer-to-core distance are plotted in Fig. 6(b). Fitting the experimental results allows us to come up with the empirical relation as $\Delta\lambda_c(y) = 2.823e^{-5}y^3 - 0.001y^2 + 0.002y + 0.207$. Similar to Section 3.1, the exact quantitative relationship here also depends on the set of the other parameters. The nonlinear relationship might be caused by the complex process of stress counteractions and elastic deformations inside the fiber [18,19].

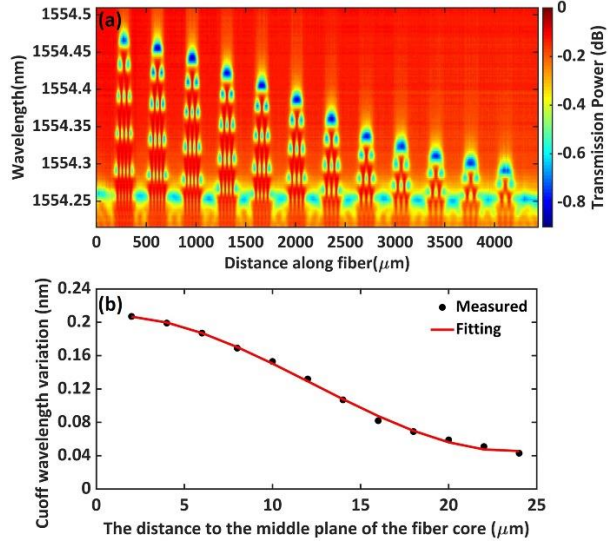

Fig.6 (a) Spectrogram of SNAP microresonators fabricated by changing the layer-to-core distance varied from 2 μm to 24 μm with the step of 2 μm from left to right, respectively. (b) Empirical relationship between the CWV and the layer-to-core distance.

In addition to the distance from the inscribed layer to the fiber core, the layer number also plays a key role in controlling the ERV. Intuitively, more layers could induce larger ERV. In order to verify the effectiveness of the multiple layer inscription for the enhancement of the ERV, we fabricated 9 SNAP microresonators with the same parameters as those used in Fig.6(a), but with different number of inscribed layers $N_L$ along the $y$ axis. The number of the inscribed layers $N_L$ is varied from 1 to 9, while the layer spacing is fixed at $\Delta y = 6 \mu m$. Notably, the SNAP microresonators with odd number of the inscribed layers were fabricated in both the central plane and symmetric layers in the cladding area, while those with even number were only fabricated in the symmetric layers of the cladding area. The spectrogram of the SNAP microresonators is shown in Fig.7(a). One can get that the CWVs of these SNAP microresonators are 0.096 nm, 0.191 nm, 0.298 nm, 0.334 nm, 0.452 nm, 0.43 nm, 0.558 nm, 0.489 nm, 0.614 nm, from left to right, respectively, We can see that the CWV increases with the increasing of $N_L$. However, the CWV of the 6$^{th}$ SNAP microresonator (from left in Fig.7) fabricated with six symmetric layers in the fiber cladding is smaller than that of the 5$^{th}$ SNAP microresonator fabricated with four symmetric layers in the fiber cladding and one layer in the central of fiber core. This's because the 6$^{th}$ SNAP microresonator contains two symmetrically inscribed layers with the layer-to-core distance of 18 μm. The CWV induced by these two layers is ~ 0.07 nm (see Fig. 6(b)), which is smaller than that inscribed in the central plane of fiber core with ~ 0.096 nm (see the 1$^{st}$ SNAP microresonator in Fig.7(a)). Similarly, the CWV of the 8$^{th}$ SNAP microresonator (from left in Fig.7) fabricated with eight symmetric layers in the fiber cladding is smaller than that of the 7$^{th}$ SNAP microresonator fabricated with six symmetric layers in the fiber cladding and one layer in the central plane of fiber core. Since the 8$^{th}$ SNAP microresonator contains two symmetrically inscribed layers with the layer-to-core distance of 24 μm. The CWV induced by these two layers is ~ 0.043 nm (see Fig. 6(b)), which is also smaller than that inscribed in the central plane of fiber core with ~ 0.096 nm. In general, increasing the number of the inscribed layers could help the fabrication of the SNAP microresonators with higher CWV and enrich a new dimensionality of axially inscription in optical fiber.

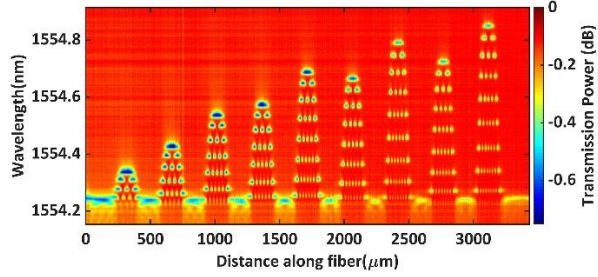

Fig.7 (a) Spectrogram of SNAP microresonators fabricated with different number of inscribed layers varied from 1 to 9 (from left to right) with the fixed layer spacing of 6 $\mu$m.

### 3.5 The repetition rate and the translating speed

For fs laser inscription, we know that the net fluence [20,21] is a key measure of exposure for process optimization, which is proportionally affected by the ratio of the repetition rate $f$ and the translating speed $v$ under a certain single pulse energy. For convenience, we define the ratio as $\upsilon = f/v$. In our SNAP fabrication, the ratio $\upsilon$ also affects the inscription process and thus the ERV. In order to investigate the influence of the ratio $\upsilon$ on the fabrication precision and robustness, we fabricated 7 SNAP microresonators with the repetition rate $f$ varied from 200 kHz to 25 kHz and the translating speed $v$ varied from 20 $\mu$m/s to 2.5 $\mu$m/s, which ensures the fixed ratio $\upsilon$ of $10^4 \mu m^{-1}$. The other fabrication parameters ($L_z, E_s, N, \Delta x, N_L, \Delta y$) are set as (150 $\mu$m, 52 nJ, 10, 3 $\mu$m, 3, 6 $\mu$m). The spectrogram of the SNAP microresonators is shown in Fig. 8(a). The CWVs of these SNAP microresonators are 0.272 nm, 0.275 nm, 0.277 nm, 0.272 nm, 0.277 nm, 0.273 nm, 0.274 nm, respectively, which are shown in Fig. 8(b). We can get that the maximal difference of CWV is ~ 5 pm, which corresponds to the ERV difference of ~ 1 $\overset{\circ}{A}$. Such a small deviation indicates excellent robustness and ultra-high fabrication precision by maintaining the ratio of the repetition rate and translating speed.

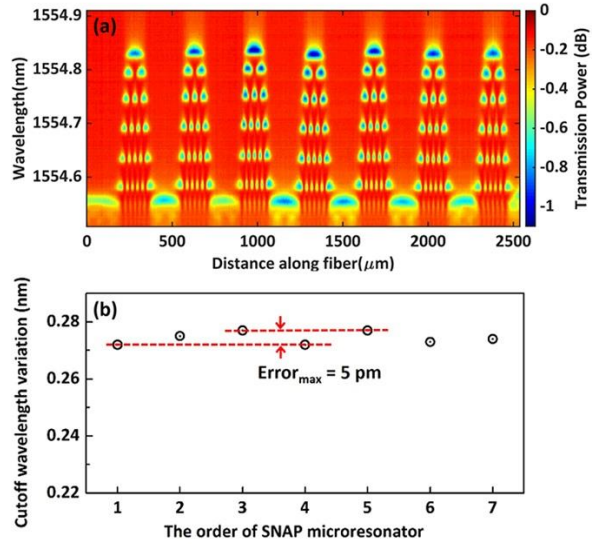

Fig.8 (a) Spectrogram of SNAP microresonators fabricated by the fixed ratio $\upsilon$ of $10^4 \mu m^{-1}$ with the repetition rate varied from 200 kHz to 25 kHz and the translating speed varied from 20 $\mu$m/s to 2.5 $\mu$m/s. (b) The CWVs of these SNAP microresonators in (a).

*3.6 The fabrication of SNAP microresonator with miniature axial size and maximal ERV*

Based on the results of the above sections, as a proof-of-principle, we optimized the variable fabrication parameters to fabricate a SNAP microresonator with both small axial size and maximal ERV. To avoid the additional loss caused by the oil ablation outside the optical fiber when the inscribed regions are too close to the fiber surface, the fabrication parameters $(E_s, N, \Delta x, N_L, \Delta y, f, v)$ were set as (89 nJ, 10, 5 $\mu$m, 5, 6 $\mu$m, 200 kHz, 10 $\mu$m/s). Moreover, we tried to set the axial length of the inscribed lines as short as possible to fabricate the SNAP microresonator with minimal axial size. Here, as an example, we set the axial inscription length $L_z$ to be 50 $\mu$m. The spectrogram of the SNAP microresonator is shown in Fig. 9(a). We can obtain that the CWV is ~ 1 nm, corresponding to the ERV of ~ 25 nm. The characteristic axial dimension of the fundamental axial mode is ~ 50 $\mu$m while the overall axial length of the structure is slightly larger than 50 $\mu$m, which is caused by the extending of the stress around the inscription regions. Besides, we calculated the intrinsic Q factor of the microresonator by measuring the FWHM of the axial eigenmode $q = 4$ at the axial position $z$ of 116 $\mu$m in Fig. 9(a). The spectrum in Fig. 9(b) shows that the FWHM is ~1.6 pm, which corresponds to the intrinsic Q factor ~ $10^6$. Overall, based on our extensive experiments and fabrication system configuration, the optimized fabrication parameters demonstrated in this section enables one to fabricate the ultra-low loss and high precision SNAP microresonators with both small axial size and maximal ERV.

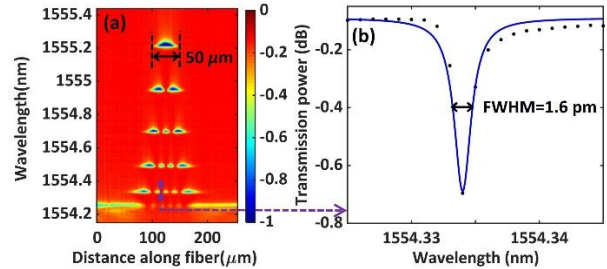

Fig.9 (a) Spectrogram of the SNAP microresonator fabricated with the axial inscribed line length of 50 $\mu$m. (b) The FWHM of the resonant mode taken from the axial mode q = 4 at the position z = 116 $\mu$m in (a).

## 4. Conclusion

In conclusion, we have experimentally investigated the relationships between the ERV of the SNAP microresonators and the controllable fabrication parameters of the in-fiber axial inscription approach with a fs laser. Using the control variate method, our investigation reveals the below general principles for the SNAP fabrication: (i) The ERV of the SNAP microresonator increases linearly with the number of the inscribed lines in the same inscribed layer; (ii) The ERV also increases linearly with the laser pulse energy when the spacing of the adjacent axially inscribed lines is larger than the inscribed line width; (iii) The number of the inscribed lines and the line spacing are limited by the fiber radius to avoid introducing additional losses with the inscribed regions too close to the fiber surface; (iv) The introduced ERV in the symmetric layers of the cladding area decreases versus the increasing of the layer-to-core distance. Meanwhile, increasing the number of inscribed layers is helpful to improve the ERV; (v) Fixing the ratio of the repetition rate and translating speed can lead to the same ERV, which allow to set the repetition rate and the translating speed with a certain degree of flexibility. Furthermore, by optimizing various fabrication parameters, we realized a SNAP microresonator with the characteristics of both small axial dimension and maximal ERV. Our experiments demonstrated that the axially oriented inscription approach could control the CWV of a SNAP microresonator varied from 0.01 nm to more than 1 nm with ultra-low loss, sub-angstrom precision and great robustness. Our work makes significant advancement of the SNAP fabrication platform, and also likely leads to deeper insight into the physics of fs laser inscription, including local stress phenomena.

## Funding

## Disclosures

The authors declare no conflicts of interest.